\documentclass[11pt]{article}
\def\Journal#1#2#3#4{{#1} {\bf #2} (#3) #4}
\def\GaC{\em Gravitation and Cosmology}
\def\GaCS{{\em Gravitation and Cosmology} Supplement}
\def\PAN{\em Phys.Atom.Nucl.}

\def\PLB{{\em Phys. Lett.}  B}

\def\PRD{{\em Phys. Rev.} D}

\def\JETPL{\em JETP Lett.}
\usepackage{graphicx}

\usepackage{amsmath}
\usepackage{amsxtra}
\usepackage{amstext}
\usepackage{amssymb}
\usepackage{latexsym}

\newcommand{\be}{\begin{equation}}
\newcommand{\ee}{\end{equation}}
\newcommand{\bn}{\begin{eqnarray}}
\newcommand{\en}{\end{eqnarray}}

\begin{document}

\title{On the problem of quark-lepton families}

\author{Maxim Yu. Khlopov\\Centre for Cosmoparticle Physics "Cosmion" and\\
National Research Nuclear University "Moscow Engineering Physics Institute",\\ 115409 Moscow, Russia \\
    APC laboratory \\10, rue Alice Domon et L\'eonie Duquet\\ 75205
Paris Cedex 13, France\\ khlopov@apc.univ-paris7.fr}
%
%\affiliation{}

\maketitle

\begin{abstract}

The problem of quark-lepton families is discussed in the "bottom-up" phenomenological approach to the extensions of the Standard model. It provides the possibility of the {\it Horizontal unification} of the three known families on the basis of horizontal gauge flavor $SU(3)_H$ symmetry. The new generations of quarks and leptons can exist.
If unstable and mixed with light fermions, they should contribute the CKM matrix. If stable and decoupled from known families, new generations can provide new candidates for dark matter.
%More detailed analyses of mass matrices
%with numerical results are in preparation~\cite{albinonorma}.

\end{abstract}

%\keywords{Unifying theories, Origin of families, The fourth family, The new stable family,
%Fermion masses and mixing matrices, Flavour symmetry, Majorana masses}

%\pacs{14.60.Pq, 12.15.Ff, 12.60.-i}
%\maketitle

%
\section{Introduction }
\label{introduction}
The origin of families and description of their properties should have fundamental theoretical basis. This description should reproduce the observed properties of the three known families and make definite predictions for the expected properties of new generations. Here we discuss some qualitative features of the approach of phenomenological bottom-up models based on the extension of the symmetry of the Standard model by the additional gauge group $SU(3)_H$ of the family symmetry. This review may be useful for the distinction of the "{\it spin-charge-family-theory}" \cite{norma} from other approaches to the problem of quark lepton families. On the other hand the experience of phenomenological description of flavor symmetry can be useful for further development of this theory.

The existence and observed properties of the three known quark-lepton families appeal to the broken $SU(3)_H$ family symmetry \cite{SU3}, which should be involved in the extension of the Standard model. It provides the possibility of the {\it Horizontal unification} in the "bottom-up" approach to the unified theory \cite{HU}. Even in its minimal implementation the model of {\it Horizontal unification} can reproduce the main necessary elements of the modern cosmology. It provides the physical mechanisms for inflation and baryosynthesis as well as it offers unified description of candidates for Cold, Warm, Hot and Unstable Dark Matter. Methods of cosmoparticle physics \cite{book,newbook} have provided the complete test of this model.

The extension of the Standard model also involve new generations. Stable new generations are of special interest for cosmological consequences, since they can provide candidates for the dark matter.

Here we discuss the possibilities to link physical basis of modern cosmology to the parameters of broken family symmetry, as well as the possible physical basis and properties of fourth generation.

\section{Horizontal unification}
\label{Khlopov}

Model of {\it Horizontal unification} \cite{HU} is based on the development of gauge $SU(3)_H$ flavor model of quark-lepton families \cite{SU3} and occupies a special place in the phenomenological description of known families (see \cite{book,newbook} for review and references).

\subsection{Horizontal hierarchy}
The approach \cite{SU3} (and its revival in \cite{bai}) follows the concept of local gauge symmetry $SU(3)_H$, first proposed by Chkareuli \cite{jon}. Under the action of this symmetry the left-handed quarks and leptons transform as $SU(3)_H$ triplets and the right-handed – as antitriplets. Their mass term transforms as
$3\bigotimes3=6\bigotimes \bar3$ and, therefore, can only form as a result of horizontal symmetry breaking.

This approach can be trivially extended to the case of $n$ generations, assuming the proper $SU(n)$ symmetry. For three generations, the choice of horizontal symmetry $SU(3)_H$ is the only possible choice because the orthogonal and vector-like gauge groups can not provide different representations for the left- and right-handed fermion states.

In the considered approach, the hypothesis that the structure of the mass matrix is determined by the structure of horizontal symmetry breaking, i.e., the structure of the vacuum expectation values of horizontal scalars carrying the $SU(3)_H$ breaking is justified.

The mass hierarchy between generations is related to the hypothesis of a hierarchy of such symmetry breaking. This hypothesis is called - the hypothesis of horizontal hierarchy (HHH) \cite{zurab}.

The model is based on the gauge $SU(3)_H$ flavor symmetry, which is additional to the symmetry of the Standard model. It means that there exist 8 heavy horizontal gauge bosons and there are three multiplets of heavy Higgs fields $\xi_{ij}^{(n)}$ ($i$,$j$ - family indexes,$n=1,2,3$) in nontrivial (sextet or triplet) representations of $SU(3)_H$. These heavy Higgs bosons are singlets relative to electroweak symmetry and don't have Yukawa couplings with ordinary light fermions. They have direct coupling to heavy fermions. The latter are singlets relative to electroweak symmetry. Ordinary Higgs $\phi$ of the Standard model is singlet relative to $SU(3)_H$. It couples left-handed light fermions $f_L^i$ to their heavy right-handed partners $F_R^i$, which are coupled by heavy Higgses $\xi_{ij}$ with heavy left handed states $F_L^j$. Heavy left-handed states $F_L^j$ are coupled to right handed light states $f_R^j$ by a singlet scalar Higgs field $\eta$, which is singlet both relative to $SU(3)_H$ and electroweak group of symmetry. The described succession of transitions realizes Dirac see-saw mechanism, which reproduces the mass matrix $m_{ij}$ of ordinary light quarks and charged leptons $f$ due to mixing with their heavy partners $F$. It fixes the ratio of vacuum expectation values of heavy Higgs fields, leaving their absolute value as the only main free parameter, which is determined from analysis of physical, astrophysical and cosmological consequences.

The $SU(3)_H$ flavor symmetry should be chiral to eliminate the flavor symmetric mass term. The condition of absence of anomalies implies heavy partners of light neutrinos, and the latter acquire mass by Majorana see-saw mechanism. The natural absence in the heavy Higgs potentials of triple couplings, which do not appear as radiative effects of any other (gauge or Yukawa) interaction, supports additional global U(1) symmetry, which can be associated with Peccei-Quinn symmetry and whose breaking results in the Nambu-Goldstone scalar filed, which shares the properties of axion, Majoron and singlet familon.
\subsection{Horizontal unification}
The model provides complete test (in which its simplest implementation is already ruled out) in a combination of laboratory tests and analysis of cosmological and astrophysical effects. The latter include the study of the effect of radiation of axions on the processes of stellar evolution, the study of the impact of the effects of primordial axion fields and massive unstable neutrino on the dynamics of formation of the large-scale structure of the Universe, as well as analysis of the mechanisms of inflation and baryosynthesis based on the physics of the hidden sector of the model.

The model results in physically self-consistent inflationary scenarios with dark matter in the baryon-asymmetric Universe. In these scenarios, all steps of the cosmological evolution correspond quantitatively to the parameters of particle theory.
The physics of the inflaton corresponds to the Dirac 'see-saw' mechanism of generation of the mass of the quarks and charged leptons, leptogenesis  of baryon asymmetry is based on the physics of Majorana neutrino masses. The parameters of axion CDM, as well as the masses and lifetimes of neutrinos correspond to the hierarchy of breaking of the $SU(3)_H$ symmetry of families.
\section{New generations}
\subsection{The problem of New generations}
Modern precision data
on the parameters of the Standard model do not exclude \cite{Maltoni:1999ta} the existence of
the  4th generation of quarks and leptons. Even more new generations are possible, if their contribution to these parameters is negligible, e.g. due to decoupling of very heavy quarks and leptons of these new generations.

If the 4th generation is mixed with the three known families, quarks and leptons of this generation are unstable, as it is the case for the current implementation of the "{\it spin-charge-family-theory}". Then the fermions of such 4th sequential generation should contribute in the matrix of quark mixing and their effect should be observed as violation of orthogonality of the Cabibbo-Kobayashi-Maskava matrix for three known generations. Therefore such violation would favor the existence of the 4th sequential family \cite{KM4}.

The problem of 4th sequential generation is related with the 4th neutrino, which should be heavier than $m_Z/2$ ($m_Z$ is the mass of Z boson), what follows from the Z boson width. There should be some fundamental explanation for such a great difference in mass of this neutrino and neutrinos of the three light families. In the case of stable 4th generation this fact finds natural explanation in a new conserved charge, which 4th generation possess.
\subsection{Stable 4th generation}

The hypothesis of stable 4th generation was connected in \cite{Belotsky} with the phenomenology of superstrings. In this phenomenology the GUT symmetry has a rank higher than the rank of the symmetry of the standard model. On the other hand, the Euler characteristic of the topology of the compactified six dimensions defines in this approach the number of generations of quarks and leptons, which can be both 3 and 4. The difference in the ranks of the symmetry groups of grand unification and the standard model implies the existence of at least one new conserved charge, which may be associated with quarks and leptons of the fourth generation. This may explain the stability of the lightest quarks and leptons (massive neutrinos) of the 4th generation and provides the basis for composite dark matter model. The latter is discussed in the contribution \cite{shibaev,dadm} to these proceedings.

\section{Conclusions}
\label{conclusions}
In its current implementation the qualitatively important features of "{\it spin-charge-family-theory}" are related with the principal existence of the 4th sequential family, which should be mixed with the 3 known families. If decoupling of the 4th family from the three known families is possible, the known families can be considered in the framework of $SU(3)_H$ flavor symmetry and the experience, gained in the development of the model of horizontal unification (MHU) will be useful. On the other hand, MHU offers guidelines, following which development of "{\it spin-charge-family-theory}" can give physical mechanisms of inflation, baryosynthesis and proper candidates for dark matter.

\appendix*

\end{document}